\documentclass[prd,nofootinbib,notitlepage,10pt,aps]{revtex4-1}
\usepackage{amsmath,amssymb,slashed}
\usepackage[utf8]{inputenc}
\usepackage{graphicx}
\usepackage{xcolor}
\usepackage{hyperref}
\usepackage{bm,bbold}

\def\!{\mskip-\thinmuskip}

\newcommand{\bra}[1]{\langle #1|}
\newcommand{\ket}[1]{|#1\rangle}

\newcommand{\mean}[1]{\langle #1 \rangle}

\newcommand{\di}{{\rm d}}
\newcommand{\Tr}{{\rm Tr}}
\newcommand{\ii}{{\rm i}}

\def\wT{{\widehat T}}
\def\wj{{\widehat j}}

\def\wP{{\widehat P}}

\def\wW{{\widehat W}}

\def\wPi{{\widehat{\Pi}}}

\def\wrho{{\widehat{\rho}}}

\def\wA{{\widehat{A}}}
\def\wB{{\widehat{B}}}

\newcommand{\tr}{{\rm tr}}  
\newcommand{\q}{\textrm{q}}  
\newcommand{\e}{{\rm e}}

\newcommand{\betav}{\boldsymbol{\beta}}

\newcommand{\krm}{{\rm k}}

\newcommand{\Psibar}{{\overline \Psi}}

\newcommand{\be}{\begin{equation}}
\newcommand{\ee}{\end{equation}}                                                                    
\newcommand{\bea}{\begin{eqnarray}}
\newcommand{\eea}{\end{eqnarray}}

\newcommand{\ped}[1]{_{\textup{#1}}}

\newcommand{\group}[1]{\relax\ifmmode\mathsf{#1}\else\textsf{#1}\fi}	
\renewcommand{\vec}[1]{\ensuremath{\mathchoice				
                     {\mbox{\boldmath$\displaystyle\mathbf{#1}$}}
                     {\mbox{\boldmath$\textstyle\mathbf{#1}$}}
                     {\mbox{\boldmath$\scriptstyle\mathbf{#1}$}}
                     {\mbox{\boldmath$\scriptscriptstyle\mathbf{#1}$}}}}

\begin{document}

\title{Polarization as a signature of local parity violation in hot QCD matter}
\author{F. Becattini}
\affiliation{Universit\`a di Firenze and INFN Sezione di Firenze, Via G. Sansone 1, 
	I-50019 Sesto Fiorentino (Firenze), Italy}
\author{M. Buzzegoli}
\affiliation{Department of Physics and Astronomy, Iowa State University, Ames, Iowa 50011, USA}
\author{A. Palermo}
\affiliation{Universit\`a di Firenze and INFN Sezione di Firenze, Via G. Sansone 1, 
	I-50019 Sesto Fiorentino (Firenze), Italy}
\author{G. Prokhorov}
\affiliation{Joint Institute for Nuclear Research, 141980 Dubna, Russia}

\begin{abstract}
	We show that local parity violation due to chirality imbalance in relativistic nuclear 
	collisions can be revealed by measuring the projection of the polarization vector onto the
	momentum, i.e. the helicity, of final state baryons. The proposed method does not require 
	a coupling to the electromagnetic field, like in the Chiral Magnetic Effect. By using linear 
        response theory, we show that, in the presence of a chiral imbalance, the spin 1/2 baryons 
	and anti-baryons receive an additional contribution to the polarization along their momentum 
	and proportional to the axial chemical potential. The additional, parity-breaking, contribution
        to helicity can be detected by studying helicity-helicity azimuthal angular correlation.
\end{abstract}

\maketitle

\section{Introduction}

The vacuum state of the Quantum Chromodynamics (QCD) plays a crucial role in the understanding 
of strong interactions phenomenology. The study the Quark Gluon Plasma (QGP) in relativistic 
heavy ion collisions provides essential information on QCD at high temperature, but it may also 
shed light on QCD vacuum. Indeed, thanks to the high temperatures, non-trivial topological 
configurations can be produced with sufficiently high probability~\cite{McLerran:1990de} through 
a classical thermal transition process called sphaleron~\cite{Manton:1983nd}. Given the random nature
of this process, the topological charge fluctuates on an event by event basis~\cite{Kharzeev:2001ev} 
in nuclear collisions and vanishes when averaged over many events.

The local topological fluctuations are transferred to the chirality of fermions through the axial 
anomaly~\cite{Adler:1969gk,Bell1969} and an imbalance between right-handed and left-handed 
quarks, hence a local parity violation, is thereby generated \cite{Kharzeev:1998kz}. Thanks to 
the chiral symmetry of QGP, the imbalance is maintained through all the evolution of the 
plasma~\cite{Kharzeev:2007jp}. 
The asymmetry between the number of right-handed and left-handed fermions can be included in 
a hydrodynamic picture with an axial chemical potential~\cite{Kharzeev:2007jp,Fukushima:2008xe}.

Local parity violation has been investigated in heavy-ion collisions via the so-called 
Chiral Magnetic Effect (CME)~\cite{Fukushima:2008xe}. This phenomenon, experimentally found in
condensed matter, is the generation of an electric current parallel to a magnetic field and
proportional to the axial chemical potential. The CME is expected to bring about a charge-dependent 
azimuthal asymmetry in the spectrum of produced particles \cite{Abelev:2009ac}. 
However, backgrounds unrelated to the CME are difficult to evaluate~\cite{Kharzeev:2015znc,Bzdak:2012ia} 
and dedicated experiments with isobar collisions~\cite{Voloshin:2010ut,Deng:2016knn,Adam:2019fbq} 
have been proposed and are currently ongoing to finally demonstrate its existence. From the 
phenomenological standpoint, there are large uncertainties on the magnitude of the magnetic field 
in the plasma phase and this affects the quantitative assessment of the CME.

Lately, the STAR experiment at RHIC measured a global $\Lambda$ polarization \cite{STAR:2017ckg} 
which turned out to be in very good agreement with predictions based on the hydrodynamic model of the QGP 
\cite{Becattini:2020ngo}. Also, the experiments proved to be able to measure it differentially 
in momentum space \cite{Adam:2018ivw,Adam:2019srw}. These findings have opened a new window in the 
field of relativistic heavy ion physics with spin and polarization being newly available probes
to study the QGP and its properties. 

In this work, we propose to study and detect local parity violation by measuring the longitudinal 
component of polarization, that is helicity, of baryons produced in the collision, particularly 
$\Lambda$ hyperons. We will show that, if the axial chemical potential does not vanish at hadronization, the
helicity of baryons is predicted to have an additional, parity-breaking, contribution with a specific 
azimuthal dependence in the transverse momentum plane. A similar idea was put forward by the authors 
of ref.~\cite{Finch:2018ner}, who proposed to correlate net helicity of $\Lambda$'s with charge 
separation due to CME. In fact, our proposed method does not require, like in the CME, the mediation 
of the electromagnetic field and it thus allows to evade some of the related uncertainties.
\begin{figure}[thb]
    \centering
    \includegraphics[width=0.5\columnwidth]{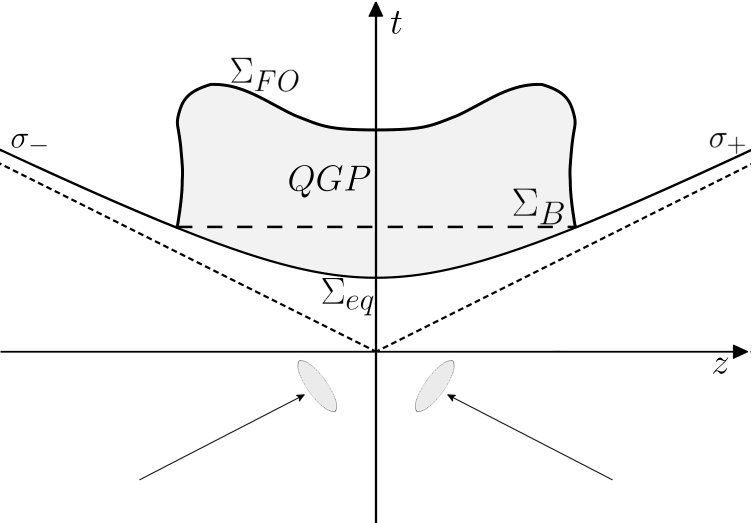}
    \caption{Space-time diagram of a relativistic nuclear collision in the
    center-of-mass frame. $\Sigma\ped{eq}$ is the 3D hypersurface where local
    thermodynamic equilibrium is achieved, $\Sigma\ped{FO}$ is the freeze-out
    hypersurface. The $\sigma_\pm$ are the side branches subsets of $\Sigma\ped{eq}$ 
    and $\Sigma\ped{B}$ is the portion of 
    hyperplane connecting the limiting surfaces of $\Sigma\ped{FO}$.}
    \label{fig:fo}
\end{figure}

\section{Polarization induced by an axial chemical potential}

The mean spin vector of a spin $1/2$ hadron in a nuclear collision can be calculated 
by using the formula~\cite{Becattini:2020sww}
\begin{equation}\label{eq:PolS}
S^\mu(p)=\frac{1}{2}\frac{\int_\Sigma \di\Sigma \cdot p\,\tr\left[\gamma^\mu\gamma^5 W_+(x,p)\right]}
{\int_\Sigma \di\Sigma \cdot  p\, \tr \left[W_+(x,p)\right]}
\end{equation}
where $\Sigma$ is the so-called freeze-out hypersurface (see fig.~\ref{fig:fo})
\footnote{Precisely, $\Sigma$ is the hypersurface including $\Sigma_{FO}$ and the
two hyperbolic branches $\sigma_+$ and $\sigma_-$} and $W_+$ is 
the future time-like part (that is the particle part) of the Wigner function:
\be\label{wigfun}
   W_+(x,p)_{AB} = \theta(p^0) \theta(p^2)
   \frac{1}{(2\pi)^4} \int \di^4 y \; \e^{-\ii p \cdot y} 
  \Tr ( \wrho :\Psibar_B (x+y/2) \Psi_A (x-y/2):)  .
\ee
Because of the integration over the hypersurface, the four-momentum $p$ argument of the Wigner 
becomes on-shell in the \eqref{eq:PolS}, that is $p^2=m^2$ \cite{Becattini:2020sww}.

In the equation \eqref{wigfun} $\wrho$ is the density operator and $:\;\;:$ denotes normal
ordering. In the hydrodynamic model of the nuclear collision, to a good approximation, 
corresponding to ideal dissipationless hydrodynamics, is the local equilibrium density operator:
\begin{equation}\label{leq}
\wrho\ped{LE} = \frac{1}{Z\ped{LE}} \exp \left[- \int_\Sigma \di \Sigma_\mu 
 \left( \wT^{\mu\nu} \beta_\nu - \sum_i \zeta_i \wj_i^\mu \right) \right],
\end{equation}
where $\beta = (1/T) u$ is the four-temperature vector and $\zeta_i = \mu_i/T$ are the
temperature-scaled chemical potentials, which are connected to the conserved currents 
$\wj_i$. In the equation \eqref{leq}
$\beta,\zeta_i$  are functions of the space-time point and may fluctuate on an 
event-by-event basis.

If there is a chiral imbalance in the QGP, the exponent in \eqref{leq} should include an additional term:
\be\label{chiral}
 \int_\Sigma \di \Sigma_\mu \; \zeta_A \wj_A^\mu, \qquad \qquad \zeta_A = \frac{\mu_A}{T}\, .
\ee
where $\wj_A$ is the axial current and $\mu_A$ the axial chemical potential at the hadronization. 
Even though the axial current is not conserved in the hadronic phase, the term \eqref{chiral} 
must be there if a chiral imbalance is generated when the plasma achieves local thermodynamic
equilibrium, what can be shown by using the Gauss theorem to work out the actual density 
operator \cite{Becattini:2019dxo} (see Appendix~\ref{append}). The term 
\eqref{chiral} may violate parity (the operator $\wrho$ does not commute with the reflection 
operator $\wPi$) if the function $\zeta_A$ has a scalar component, that is a component which 
does not change sign 
under reflection~\cite{Buzzegoli:2018wpy}. It is important to stress that this component of
$\zeta_A$ fluctuates on an event-by-event basis and averages to zero over many events, so as
to keep parity breaking local, in a single event and not global, as mentioned above.
Presently, there is quite a large uncertainty on the value of the axial chemical potential 
$\mu_A$. Several estimates have been proposed based on the early-stage glasma model
\cite{Kharzeev:2001ev,Lappi:2006fp,Jiang:2016wve} or lattice simulations~\cite{Mueller:2016ven,Mace:2016shq}
which are then used to study its evolution in the QGP with hydrodynamic
codes~\cite{Hirono:2014oda,Jiang:2016wve,Shi:2017cpu,Lin:2018nxj,Liang:2020sgr}. The calculations
in \cite{Shi:2017cpu} imply $\zeta_A = {\cal O}(10^{-2})$ at hadronization \cite{LiaoPrivate}.

Anyhow, it is expected that the term \eqref{chiral} is a ``small'' correction to the
operators in \eqref{leq} which does not affect much the shape of the momentum spectra 
(except for specific asymmetries such as those sought in the CME) and yet, it may have a
sizeable impact on the polarization of emitted hadrons. Using the linear response theory
to expand the local equilibrium operator, we determine, at the leading order, the mean spin vector 
of a free fermion induced by the axial chemical potential (see Appendix~\ref{append}):
\be\label{spinaxial}
S^\mu_{\chi}(p) \simeq \frac{g_h}{2} \frac{\int_\Sigma \di\Sigma\cdot p \; 
\zeta_A n\ped{F}\left(1-n\ped{F}\right)}{\int_\Sigma \di\Sigma \cdot p \; 
n\ped{F}} \frac{\varepsilon p^\mu- m^2\hat t^\mu}{m \varepsilon}
\ee
where $g_h = G_{A1}(0)$ is the axial charge of the baryon species, 
which depends on the  transformation properties of the axial current in flavour space.
In the equation \eqref{spinaxial} $n\ped{F}$ is a shorthand for the Fermi-Dirac distribution function:
\be\label{fermi}
n\ped{F} =  \frac{1}{\e^{\beta(x)\cdot p - \sum_i \zeta_i q_i} + 1}
\ee
and $\hat t^\mu=\delta^\mu_0$ is the unit time-like vector in the centre-of-mass frame 
(see fig.~\ref{fig:fo}). The appearance of an explicit dependence on a particular
vector such as $\hat t$ is owing to the fact that the axial charge:
$$
 \int_\Sigma \di \Sigma_\mu \wj^\mu_A
$$
is not an actual scalar quantum operator for it depends on the integration 
hypersurface \cite{Becattini:2021suc}, being the axial charge operator not divergenceless.
Indeed the vector $\hat t$ can be viewed as the average normal vector to the hypersurface
$\Sigma_{\rm FO}$ in fig.~\ref{fig:fo}.
This mean spin vector adds to  the already known contribution from hydrodynamics, 
namely the well known from vorticity \cite{Becattini:2013fla} and the recently 
found contributions from the shear tensor \cite{Becattini:2021iol,Liu:2021uhn,
Becattini:2021iol}, resulting in a total spin polarization
vector:
\be\label{spintot}
 S^\mu(p) = S^\mu_{\rm hyd}(p) + S^\mu_\chi(p)
\ee
for a set of events with given $\zeta_A$. Averaging over many events will lead to
a cancellation of all parity-breaking terms of $S_\chi(p)$, as has been emphasized.

If $\zeta = {\cal O}(10^{-2})$, the magnitude of the spin vector \eqref{spinaxial} 
is comparable to the one from hydrodynamics in the eq.~\eqref{spintot}. However, the former 
peculiarly differs from the latter in that it is just longitudinal, that is directed 
along the particle momentum. To prove it, let us back boost \eqref{spinaxial} to 
the rest frame of the particle:
\begin{equation}\label{backboost}
    \vec{S}_0=\vec{S}-\frac{\vec{p}}{\varepsilon(\varepsilon+m)}\vec{S}\cdot\vec{p},
\end{equation}
yielding:
\be\label{spinaxial-rest}
 {\bf S}_{0,\chi} = h_{\chi}({\bf p}) \hat{\bf p},
\ee
with $\hat{\bf p}=\vec{p}/|\vec{p}|$ and:
\be\label{fchi}
 h_\chi({\bf p}) = \frac{g_h}{2}\frac{|\vec{p}|}{\varepsilon}
 \frac{\int_\Sigma \di\Sigma\cdot p \; \zeta_A 
 n\ped{F}\left(1-n\ped{F}\right)}{\int_\Sigma \di\Sigma \cdot p \; n\ped{F}}.
\ee
Altogether, the axial chemical potential induces an additional contribution to the
helicity of spin 1/2 baryons \footnote{We define helicity as the scalar product 
of the momentum and the spin vector in the rest frame. However, helicity is also 
defined as the scalar product of the momentum and the spin vector in the same 
reference frame. The two definitions differ, according to the equation \eqref{backboost}
by a factor $m/\varepsilon$.}:
\be\label{heli}
  \vec{S}_{0,\chi} \cdot \hat{\bf p} = h_\chi({\bf p}).
\ee
which applies to anti-baryons as well being the axial current invariant by charge 
conjugation. 

Since $h_\chi$ depends on an axial chemical potential which fluctuates event-by-event
with zero mean, it vanishes when averaged over many events. Therefore, the term 
\eqref{spinaxial-rest} does not contribute to the overall mean spin vector measured 
by the experiments. Notwithstanding, this fluctuating contribution can be detected,
what will be proposed in the next Sections.

\section{Helicity and symmetry of a nuclear collision}

\begin{figure}
    \centering
    \includegraphics[width=0.6\columnwidth]{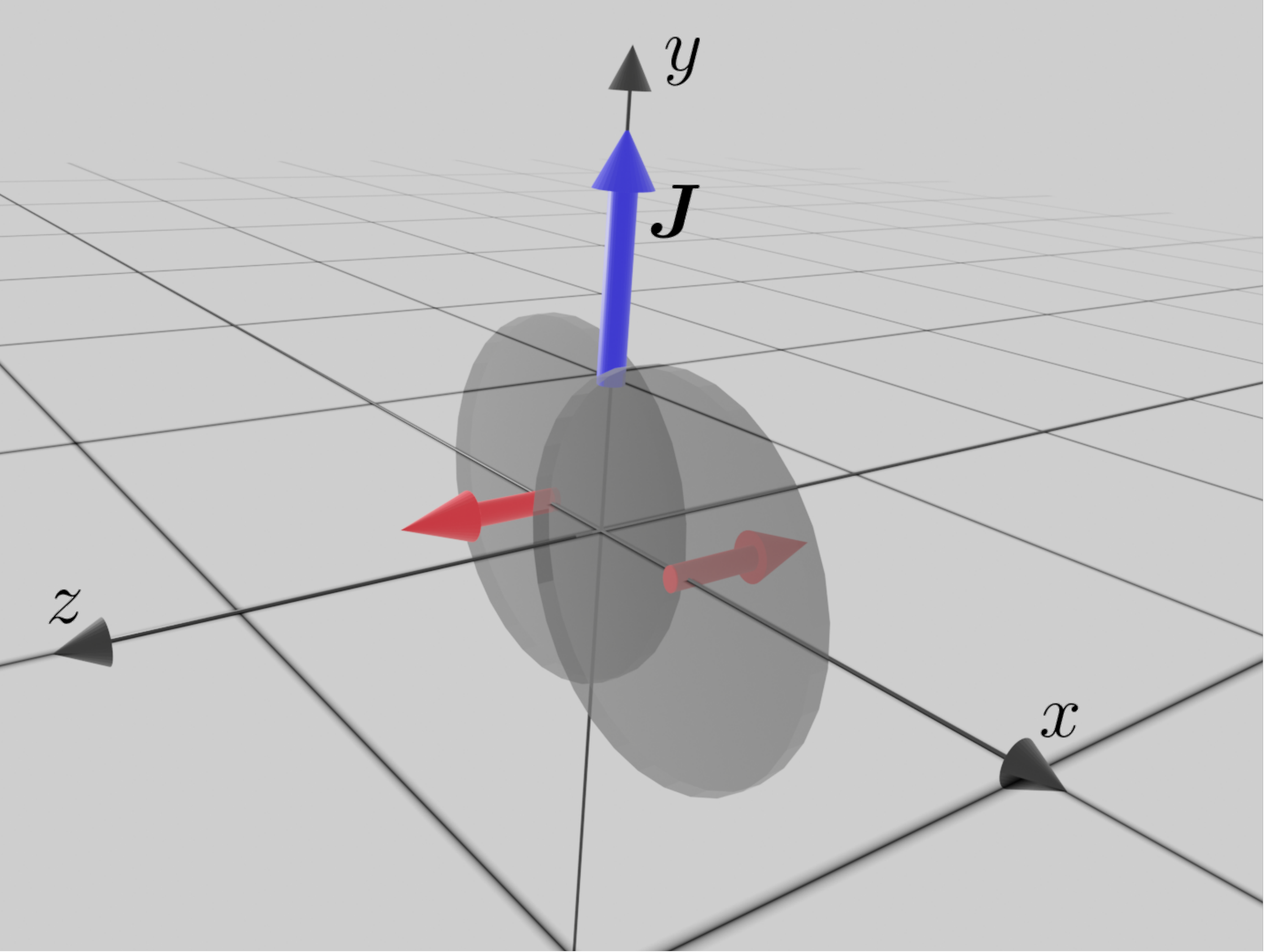}
    \caption{(Color online) Geometry of a relativistic heavy ion collision. The system is symmetric
    by rotation around $\vec{J}$ by an angle $\pi$ and is invariant by reflection with
    respect to the reaction plane ($zx$ plane). Combining the two symmetries, the system
    is invariant by total reflection.}
    \label{refframe}
\end{figure}

The average high energy nuclear collision has two remarkable geometrical symmetries: parity $\Pi$ 
and rotation of an angle $\pi$ around the angular momentum direction $R_J(\pi)$ (see fig.~\ref{refframe}). 
These geometrical symmetries should be reflected into the shape of the freeze-out hypersurface and 
the properties of the density operator and its local equilibrium approximation, that is eq.~\eqref{leq}. 
Indeed, the operator commutes with the quantum operators corresponding to $\Pi$ and $R_J(\pi)$, which 
implies that the fields $\beta$ and $\zeta_i$ should fulfill those symmetries as well. For instance, the 
four-temperature $\beta$ fulfills these relations under reflection:
$$
    \beta^0(x^0,-{\bf x}) = \beta^0(x^0,{\bf x}), \qquad 
    \betav (x^0,-{\bf x}) = -\betav(x^0,{\bf x}).
$$
On the other hand, as has been mentioned, a local parity breaking occurs if the axial chemical 
potential in a single collision event does not behave as a pseudo-scalar function, that is if:
$$
    \zeta_A(x^0,-{\bf x}) \ne - \zeta_A(x^0,{\bf x})
$$
while rotational symmetry $R_J(\pi)$ is supposedly preserved\footnote{Note that the freeze-out
hypersurface can be parametrized as $x^0=f({\bf x})$ and the function $f({\bf x})$ must
be parity-invariant, so that the argument $x^0$ does not change by reflection if the 
function $\zeta_A$ is restricted to the freeze-out hypersurface.}.

These geometrical symmetries, or lack thereof, have an exact match in momentum space (see discussion 
in ref.~\cite{Becattini:2017gcx}). Particularly, if parity is conserved, momentum spectra must be 
invariant by reflecting ${\bf p} \to -{\bf p}$. Likewise, the mean spin vector, being a pseudo-vector,
should fulfill:
$$
  {\bf S}_0 (-{\bf p})= {\bf S}_0 ({\bf p})
$$
and helicity should be a pseudo-scalar in momentum space. On the other hand, if parity is broken, helicity 
can acquire a scalar component in momentum space. This is most easily seen in the simple case
of a constant $\zeta_A$ over the freeze-out hypersurface, which turns the \eqref{heli} in the 
very simple and suggestive:
$$
  h_\chi({\bf p}) =\frac{g_h}{2}\frac{|\vec{p}|}{\varepsilon}\zeta_A 
$$
under the approximation of $1 - n_F \sim 1$ in the \eqref{fchi}. 
In general, one can expand the function $\zeta_A(x)$ at the freeze-out into multipolar
components, thus separating the parity-conserving (odd $l$) from the parity-breaking (even $l$) terms:
\be\label{multipol}
  \zeta_A(x) = \sum_{l=0}^\infty \sum_{m=-l}^l Z_{lm}(r) Y^l_m(\theta,\varphi)\, ,
\ee
where $Y^l_m$ are the spherical harmonics. Correspondingly, the helicity function has a multipolar 
expansion in momentum space:
\be\label{multipol2}
  h_\chi({\bf p}) = \sum_{l=0}^\infty \sum_{m=-l}^l H_{lm}(p) Y^l_m(\theta_{\bf p},\phi_{\bf p})
\ee
with parity-conserving odd $l$ terms and parity-breaking even $l$ terms.
Note, however, that the relations between the $H_{lm}$ and the $Z_{lm}$
are not straightforward because of the non-trivial dependence on the coordinates 
of the Fermi-Dirac distribution in the equation ~\eqref{fchi}. Particularly, a 
coefficient $Z_{lm}$ in the eq.~\eqref{multipol} cannot be reconstructed from the 
measurement of one coefficient $H_{lm}$ with the same couple of integers. In fact, 
many integers $(l,m)$ of $H_{lm}$ can contribute to one multipolar coefficient $Z_{LM}$
and vice-versa.

\section{Parity violation and helicity azimuthal dependence}

Local parity violation in the helicity spectrum can be established, in a model independent way,
by studying the azimuthal dependence of, e.g. $\Lambda$ hyperon helicity in the transverse 
plane to verify the non-vanishing even $l$ terms in the expansion \eqref{multipol2}. 
Let us consider, for simplicity, particles emitted at midrapidity in a heavy ion collisions, i.e. with
vanishing longitudinal momentum $p_z=0$; the momentum vector ${\bf p}$ is then only transverse
and can be described by a magnitude $p_T$ and the azimuthal angle $\phi$ with respect to the 
reaction plane $y=0$ in figure~\ref{refframe}. In this case, the expansion \eqref{multipol2} 
becomes a single-variable Fourier expansion in the azimuthal angle $\phi$. The helicity function 
can be split into a parity preserving pseudo-scalar part $h_P$ and a parity breaking scalar part $h_S$.
Taking into account the rotational symmetry $\phi \to \pi - \phi$ and their transformation
properties under reflection $\phi \to \pi + \phi$, they can be written as:
\begin{align}\label{azimuth}
  h_P(p_T,\phi) &= \sum_k P_k(p_T) \sin [(2k+1) \phi], \\ \nonumber
  h_S(p_T,\phi) &= \sum_k S_k(p_T) \cos [2k \phi].  
\end{align}
The above forms are dictated by symmetry, hence they are completely general and model-independent.
The models, amongst which the local equilibrium model with axial chemical potential, 
in principle predict the function \eqref{fchi} and, consequently, the momentum dependent 
coefficients of $P_k$ and $S_k$ in the \eqref{azimuth}. 

The hydrodynamic polarization in eq.\eqref{spintot} does not break parity and does not 
contribute to $h_S$, but only to $h_P$. As we have emphasized, unlike for the $P_k$'s, 
the $S_k$'s average to zero over many events and suitable observables 
must be devised to detect them. For instance, by retaining only the leading harmonics in the 
\eqref{azimuth}, the helicity squared reads:
\be\label{helsquare}
 h^2({\bf p}_T) = (S_0 + P_0 \sin \phi)^2 = S_0^2 + P_0^2 \sin^2 \phi + 2 S_0 P_0 \sin \phi
\ee
and, assuming that $S_0$ and $P_0$ are uncorrelated, being $\langle\langle S_0 \rangle\rangle = 0$ 
when averaging over many events, one has:
\be\label{helsquare2}
\langle\langle h^2({\bf p}_T) \rangle\rangle = 
\langle\langle S_0^2 \rangle\rangle + \langle\langle P_0^2 \rangle\rangle \sin^2 \phi\, .
\ee
The constant term $\langle\langle S_0^2 \rangle\rangle$ is non-vanishing and, at least
in principle, one could think of measuring it by fitting the $h^2(\phi)$ azimuthal 
function. However, since helicity can only be measured through the fluctuating 
angle between the momentum of the $\Lambda$ and the momentum of the decay proton in 
the $\Lambda$ rest frame, it would be hard to disentangle a mean value of the helicity 
squared from the fluctuation variance. Moreover, an accurate identification of the 
reaction plane is needed (not its orientation though) which might be 
difficult to achieve. 
\begin{figure}[t]
    \centering
    \includegraphics[width=0.6\columnwidth]{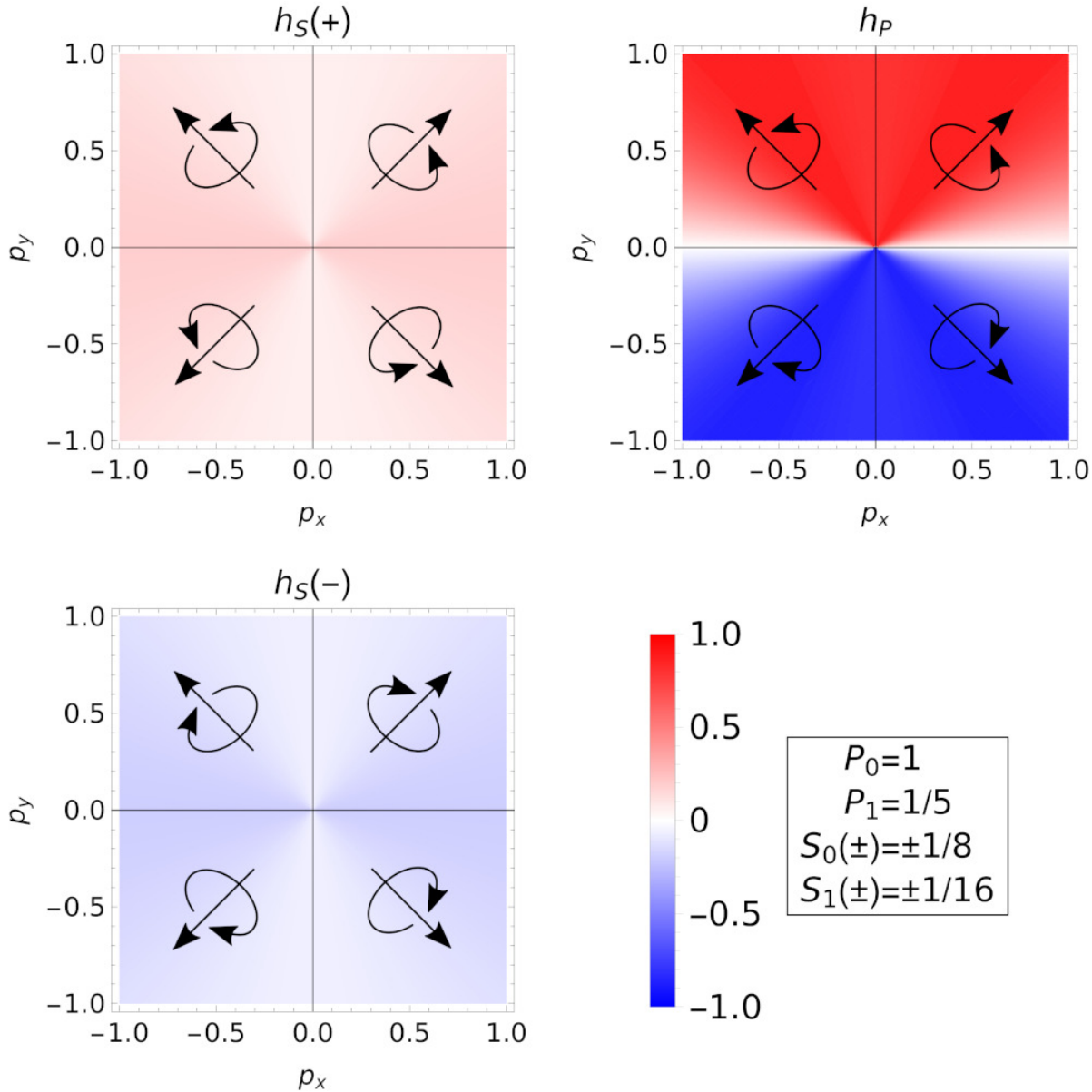}
    \caption{(Color online) Examples of the distributions of the scalar, parity-breaking, component of 
    the helicity (left) and of the pseudoscalar component (right) in the transverse momentum 
    plane. The contour plots show the profile of the helicity calculated with the 
    Fourier expansion \eqref{azimuth} and parameter values quoted in the right bottom 
    corner. The parity-breaking component fluctuates on an event-by event basis with 
    positive or negative values (left).}
    \label{fig:HContour}
\end{figure}

A better and definitely more realistic method is based on the measurement of the 
helicity-helicity angular correlation in the same event. Azimuthal polarization 
correlations have been proposed to detect the vortical structure of the hydrodynamic
motion \cite{Pang:2016igs} and we find here that they can be used to detect the
chirality imbalance as well. Suppose that two (or more) 
hyperons are emitted in the same event at two different angles $\phi$ and $\phi+\Delta \phi$ 
and also suppose, for illustrative purpose, that there is no sizeable spin-spin 
two-particle correlation. Then, if
$$
  n({\bf p}_{T1},{\bf p}_{T2}) = \frac{\di N}{d^2 {\bf p}_{T1} \di^2 {\bf p}_{T2}}
$$  
is the two-particle momentum spectrum, and $N$ its integral, we have:
\be\label{helcorr}
 \langle h_1 h_2(\Delta \phi) \rangle = \frac{1}{N} \int \di^2 {\bf p}_{T1} \di^2 {\bf p}_{T2} 
 \delta(\phi_2-\phi_1-\Delta\phi) \, 
 h_1({\bf p}_{T1}) h_2({\bf p}_{T2}) n({\bf p}_{T1},{\bf p}_{T2})
\ee
which is expected to receive contributions from the parity violating terms. Neglecting
momentum correlations and the azimuthal anisotropies of the spectrum, such as elliptic flow, 
which introduce just small corrections, and retaining only the leading harmonics just like in 
equation \eqref{helsquare}, one has:
\begin{align*}
 \langle h_1 h_2(\Delta \phi) \rangle & \simeq \frac{1}{2\pi} 
 \int_0^{2\pi}\!\! \di \phi \left( \bar S_0^2 + \bar P_0^2 \sin^2 \phi \cos \Delta\phi\right) \\
 &= \bar S_0^2 + \frac{1}{2} \bar P_0^2 \cos \Delta\phi,
 \end{align*}
where the bar stands for transverse momentum average. The first term now survives 
the averaging over many events, so that a pedestal in the helicity-helicity azimuthal 
correlation function, like in eq.~\eqref{helsquare2}, signals a local parity violation. 
The constant, parity-breaking term, $S_0$ can be highlighted by integrating the 
equation \eqref{helcorr} in $\Delta \phi$; it can be readily shown that, if momentum 
correlations are negligible as it was supposed for the equation \eqref{helcorr}:
$$
  \frac{1}{2\pi}  \int_0^{2\pi}\!\! \di \Delta \phi \; 
  \langle h_1 h_2(\Delta \phi) \rangle = \bar S_0^2
$$ 
It is important to stress that the correlation function \eqref{helcorr}, as well as 
other possible combinations of two helicities, does not require the identification 
of the reaction plane and can be measured by means of the angles between 
the $\Lambda$ momentum and the proton momentum in the $\Lambda$ rest frame.

While a non-vanishing value of the $\bar S_0^2$ is a clear signal of
parity violation, one may wonders whether parity violation can be generated only
by a genuine hot QCD-generated axial imbalance. Indeed, for the case of $\Lambda$,
a possible source of background is the parity-violating polarization transfer in 
the weak $\Xi \to \Lambda$ decay. A quantitative assessment certainly goes beyond the
scope of the present work; we just remark that secondary $\Lambda$s from $\Xi$
decays can be selected out through the displacement of their production point from
the primary vertex of the collision, what makes this background not irreducible.

\section{Conclusions and outlook}

To summarize, we have shown that the spin polarization vector of hyperons can be 
used to reveal local parity violation in hot QCD matter in relativistic
heavy ion collisions. The helicity of $\Lambda$s acquires a term which is proportional 
to the fluctuating parity-breaking axial chemical potential, that we calculated in the linear 
approximation. To detect this contribution, we propose to measure the angular 
azimuthal correlation of the helicity of $\Lambda$ pairs in the same event through 
the measurement of the angle between the momentum of the hyperon and the momentum 
of the decay proton in its rest frame. For this purpose, a full quantitative study 
of the relation between the axial chemical potential distribution and the corresponding 
helicity pattern would be an important point of a future analysis. 

\section*{Acknowledgments}
We are grateful to J. F. Liao and M. Lisa for very useful discussions. The work of 
GP is supported by RFBR Grant 18-02-40056. The work of MB was carried out while 
he was in Florence, supported by the fellowship {\em Polarizzazione nei fluidi 
relativistici}.


\bibliographystyle{apsrev4-1}

%

\clearpage
\appendix
\section{Calculation of the axial chemical contribution to the spin 
polarization vector}
\label{append}

In this Appendix section we provide the detailed derivation of the contribution
of the axial chemical potential to the polarization vector of a spin
$1/2$ particles in a relativistic fluid at local thermodynamic equilibrium.
We refer to the main letter for the notation.

The mean spin vector can be derived from the future time-like part of Wigner
function of the emitted particle~\cite{Becattini:2020sww}:
\begin{equation}
\label{SM_eq:PolS}
S^{\mu }(p)=\frac{1}{2}
\frac{\int _{\Sigma }\di  \Sigma \cdot p\,\tr 
 \left [\gamma ^{\mu }\gamma ^{5} W_{+}(x,p)\right ]}{\int _{\Sigma }\di  \Sigma \cdot p\, 
 \tr \left [W_{+}(x,p)\right ]},
\end{equation}
where $\Sigma $ can be approximated as the freeze-out 3D hypersurface in
{Fig.~\ref{fig:fo}}. The Wigner function involves the effective hadronic
fields, which are assumed to be free:
\begin{equation}
\label{SM-wigfun}
W_{+}(x,p)_{ab} = \frac{1}{(2\pi )^{4}} \int \di^{4} s \;
\e ^{-\ii p \cdot s} \Tr ( \wrho :\Psibar _{b} (x+s/2)
\Psi _{a} (x-s/2):) .
\end{equation}
The density operator $\wrho$ in the above equation must be fixed, in the
Heisenberg representation. Therefore, in the hydrodynamic picture of the
QCD plasma, it is assumed to be the local equilibrium density operator
specified by the initial conditions \cite{Becattini:2019dxo}, that is at
the 3D hypersurface where the plasma is supposed to achieve local thermodynamic
equilibrium ($\Sigma _{\mathrm{eq}}$ in {Fig.~\ref{fig:fo}}):
\begin{equation}
\label{truerho}
\wrho = \frac{1}{Z} \exp \left [- \int _{\Sigma _{\mathrm{eq}}} \di 
\Sigma _{\mu }\left ( \wT^{\mu \nu } \beta _{\nu }- \zeta _{A}
\wj_{A}^{\mu }\right ) \right ].
\end{equation}
For the sake of simplicity, we have neglected all terms involving the conserved
currents except for the axial current operator $\widehat{j}_{A}$ 
\footnote{In this work the axial current of the free Dirac field is defined 
as $\widehat{j}_A^\mu = \bar\Psi \gamma^\mu \gamma^5 \Psi$ with 
$\gamma^5 = {\rm diag}(I,-I)$}
is the color-singlet axial current expressed in terms of the fundamental
quark and gluon fields and includes the Chern-Simons current
$\widehat{K}^{\mu }$ from anomaly~\cite{Kharzeev:2009fn} so as to be a conserved
one in the plasma phase. The exponent can be rewritten, by using the Gauss'
theorem (see {Fig.~\ref{fig:fo}}):
\begin{equation}
\label{gauss}
\int _{\Sigma _{\mathrm{eq}}} \di  \Sigma _{\mu }\left ( \wT^{\mu
\nu } \beta _{\nu }- \zeta _{A} \wj_{A}^{\mu }\right ) = \int _{
\Sigma } \di  \Sigma _{\mu }\left ( \wT^{\mu \nu } \beta _{\nu }-
\zeta _{A} \wj_{A}^{\mu }\right ) + \int _{\Omega } \di  \Omega
\left ( \wT^{\mu \nu } \partial _{\mu }\beta _{\nu }- \wj_{A}^{\mu }\partial _{\mu }\zeta _{A} - \zeta _{A} \partial _{\mu }\wj_{A}^{\mu }\right )
\end{equation}
where $\Omega $ is the space-time region encompassed by the 3D hypersurfaces
$\Sigma _{\mathrm{eq}}$ and $\Sigma = \Sigma _{\mathrm{FO}} \cup \sigma _{\pm }$
\cite{Becattini:2019dxo}. The last term in the equation {\eqref{gauss}} is
responsible for the dissipative corrections and includes a term with the
divergence of the axial current which is quasi-vanishing in the chirally
symmetric QGP phase (broken by quark masses). In the hydrodynamic approach,
the local thermodynamic equilibrium term is dominant and one can obtain
a good approximation by neglecting the second integral on the right hand
side of {\eqref{gauss}}:
\begin{equation}
\label{SM_eq:leq}
\wrho \simeq \wrho \ped{LE} = \frac{1}{Z\ped{LE}} \exp \left [- \int _{\Sigma }
\di  \Sigma _{\mu }\left ( \wT^{\mu \nu } \beta _{\nu }-
\zeta _{A} \wj_{A}^{\mu }\right ) \right ].
\end{equation}

The eq.~{\eqref{SM-wigfun}} is indeed the mean value of the Wigner operator
at the point $x$
\begin{equation*}
\wW(x,p)_{ab}= \frac{1}{(2\pi )^{4}} \int \di^{4} s \; \e ^{-
\ii p \cdot s} :\Psibar_b (x+s/2) \Psi_a (x-s/2):
\end{equation*}
and, in the hydrodynamic limit of slowly varying $\beta (x)$ compared to
the microscopic length scales, one can Taylor expand the $\beta $ field
in~{(\ref{SM_eq:leq})} from $x$ and retain only the leading term:
\begin{equation}
\label{wigapp}
\Tr (\wrho \ped{LE} \wW (x,p)) \simeq \frac{1}{Z\ped{LE}}
\Tr \left ( \wW (x,p) \exp \left [-\beta (x)\cdot \wP +
\int _{\Sigma } \di  \Sigma _{\rho }\; \zeta _{A} \wj_{A}^{\rho }\right ] \right ),
\end{equation}
where $\wP$ is the total four-momentum. The term involving the axial
current term is supposedly small compared to the first term, hence one
can expand the exponential in the {\eqref{wigapp}} with the formula:
\begin{equation*}
\e ^{\wA+\wB}=\e ^{\wA}+\int _{0}^{1} \di z\,\e ^{z \wA}\,\wB \,\e ^{-z \wA}
\,\e ^{\wA}+\cdots ,
\end{equation*}
where:
\begin{equation*}
\wA=-\beta (x)\cdot \wP,\quad \wB =\int _{\Sigma }
\di  \Sigma _{\rho }\; \zeta_{A} \wj_{A}^{\rho }.
\end{equation*}
Therefore, the response of the thermal expectation value of Wigner operator
to the axial current term $\wB$ at local equilibrium is obtained
by the previous expansion and is given by, for the particle term:
\begin{equation}
\label{eq:Linear}
\mean{\wW_{+}(x,p)}\ped{LE}\simeq \mean{\wW_{+}(x,p)}_{\beta (x)}+\Delta W_{+}(x,p)
\end{equation}
with
\begin{equation}
\label{correlat}
\Delta W_{+}(x,p)=     \int _{\Sigma }    \di  \Sigma _{\rho }(y) \;
\zeta _{A}(y)     \int _{0}^{1}    \di  z \;
\mean{\widehat{W}_{+}(x,p)\widehat{j}_{A}^\rho (y+\ii z\beta (x) )}_{c,
\beta (x)},
\end{equation}
where the symbol $\mean{\cdots}_{\beta (x)}$ denotes thermal averages with
the density operator
\begin{equation*}
\wrho _{0} = \frac{1}{Z} \exp [-\beta (x) \cdot \wP ]
\end{equation*}
i.e. the familiar homogeneous global equilibrium density operator in the
grand-canonical ensemble. The subscript $c$ on the thermal average in~{\eqref{correlat}}
signifies the connected part of the correlator, that is, for the simplest
case of two operators:
\begin{equation*}
\mean{\widehat{O}_{1} \widehat{O}_{2}}_{c}\equiv
\mean{\widehat{O}_{1} \widehat{O}_{2}}-\mean{\widehat{O}_{1}}
\mean{ \widehat{O}_{2}}.
\end{equation*}

The color-singlet axial current operator can be decomposed on the multi-hadronic
Hilbert space basis and can be written as a combination of creation and
annihilation operators \cite{Weinberg:1995mt}:
\begin{align*}
\wj_{A}^{\mu }(x) =&       \sum _{\substack{N=0 \\ M=0}}^{\infty
}\sum _{\substack{j_{1},\dots ,j_{N} \\k_{1},\dots ,k_{M}}}     \int
  \frac{\di^{3} \q '_{1}}{2 \varepsilon '_{1}} \cdots
\int   \frac{\di^{3} \q '_{N}}{2 \varepsilon '_{N}}   \int
  \frac{\di^{3} \q _{1}}{2 \varepsilon _{1}} \cdots   \int
  \frac{\di^{3} \q _{M}}{2 \varepsilon _{M}}
\\
&\times \widehat{a}^{\dagger }_{j_{1}}(q'_{1}) \cdots \widehat{a}^{\dagger}_{j_{N}}(q'_{N}) 
\widehat{a}_{k_{1}}(q_{1}) \cdots \widehat{a}_{k_{M}}(q_{M})
J^{\mu }(q',q,x)^{j_{1},\dots ,k_{M}}
\end{align*}
where the indices $j_{l}$ and $k_{l}$ label the various hadronic species
and the spin indices of the creation and annihilation operators have been
omitted. Each function $J(p',p,x)$ can be obtained by forming suitable
multi-hadronic matrix elements. In the formula {\eqref{correlat}}, most of
the above terms vanish and the predominant contribution is given by the
term with two particles of the same species $h$ as specified by the Wigner
operator, which is made of hadronic fields. Specifically, the predominant
term reads (with spin indices):
\begin{equation}
\label{relevant}
\sum _{\sigma ,\sigma '} \int
\frac{\di^{3} \q '}{2 \varepsilon _{q'}}\int
\frac{\di^{3} \q }{2 \varepsilon _{q}} \; \widehat{a}^{\dagger }_{h}(p')_{
\sigma '} \widehat{a}_{h}(q)_{\sigma } J(q,q',x)_{\sigma ,\sigma '}^{hh}
\end{equation}
and the integrand function can be obtained by taking the following matrix
element of the axial current:
\begin{equation}
\label{jmatel}
J^{\mu }(q,q',x)_{\sigma ,\sigma '}^{hh} = \bra{0} \widehat{a}_{\sigma '}(q')
\wj_{A}^{\mu }(x) \widehat{a}_{\sigma }^{\dagger }(q) \ket{0}
= \bra{q',\sigma '} \wj_{A}^{\mu }(x) \ket{q,\sigma}
\end{equation}
where creation and annihilation operators are covariantly normalized:
\begin{equation*}
[\widehat{a}_{\sigma }(q),\widehat{a}_{\sigma '}^{\dagger }(q')]_{\pm }= 2
\varepsilon \, \delta _{\sigma \sigma '}\delta ^{3}(\vec{q}-\vec{q}').
\end{equation*}
The matrix element of the axial current on two spin 1/2 hadronic states
has a well-known form which is dictated by Poincar\'{e} symmetry and Dirac
equation:
\begin{equation}
\label{jmatel2}
\bra{q',\sigma '} \wj_{A}^{\mu }(x) \ket{q,\sigma} =
\frac{1}{(2\pi )^{3}} \; \e ^{\ii Q \cdot x} \bar{u}_{\sigma '}(q')
\left [ G_{A1}(Q^{2}) \gamma ^{\mu }\gamma ^{5} + \frac{Q^{\mu }}{2m_{h}} G_{A2}(Q^{2})
\gamma ^{5} \right ] u_{\sigma }(q)
\end{equation}
with $Q=(q'-q)$ and $u(q)$ are the spinors of the hadron normalized so
as to:
\begin{equation*}
\bar{u}_{\sigma }(k) u_{\sigma '}(k)= 2m_h \, \delta _{\sigma \sigma '},
\qquad \bar{v}_{\sigma }(k) v_{\sigma '}(k)= -2m_h \, \delta _{\sigma \sigma '}\, .
\end{equation*}
The axial form factors $G_{A1}(Q^{2})$ and $G_{A2}(Q^{2})$ depend on the
flavour-space transformation properties of the axial current
$\wj_{A}$, that is whether $\wj_{A}$ includes the strange quark
term and to what extent.

Altogether, the relevant part of the axial current operator in {\eqref{correlat}} 
is obtained by plugging the {\eqref{jmatel2}} and {\eqref{jmatel}} into the~{\eqref{relevant}}:
\begin{align}
\label{predominant}
\widehat{j}_{A}^{\rho }(y+\ii z\beta ) \rightarrow \widehat{j}_{A,h}^{\rho }(y+\ii z\beta )
= \frac{1}{(2\pi )^{3}} \sum _{\sigma , \sigma '} & \int \frac{\di^{3} \q '}{2 \varepsilon _{q'}}
\int \frac{\di^{3} \q }{2 \varepsilon _{q}} \; \widehat{a}^{\dagger }_{h}(q')_{\sigma '} \widehat{a}_{h}(q)_{\sigma } \e ^{\ii Q \cdot y - z t \cdot \beta }
\\
\nonumber
& \times \bar{u}_{\sigma '}(q') \left [ G_{A1}(Q^{2}) \gamma ^{\mu
}\gamma ^{5} + \frac{Q^{\mu }}{2m_{h}} G_{A2}(Q^{2}) \gamma ^{5} \right ] u_{\sigma }(q).
\end{align}

We are now in a position to work out the {\eqref{correlat}}. The Wigner operator
can be expanded by using the normal mode expansion of the Dirac field:
\begin{equation*}
\Psi (x)= \sum _{\sigma =1}^{2}\frac{1}{(2\pi )^{3/2}}\int
\frac{\di^{3} \krm }{2\varepsilon _{k}} \left [ u_{\sigma }(k)
\e ^{-\ii k\cdot x} \widehat{a}_{h}(k)_{\sigma }+ v_{\sigma }(k)
\e ^{\ii k\cdot x} \widehat{b}^{\dagger }_{h}(k)_{\sigma }\right ]
\end{equation*}
and retaining only the particle operators $\widehat{a}_{h}$ and
$\widehat{a}^{\dagger }_{h}$:
\begin{equation}\label{wigoppart}
\widehat{W}_{+}(x,p)_{ab}=\frac{1}{(2\pi )^{3}}\sum _{\tau ,\tau '}
\int \frac{\di^{3} \krm }{2\varepsilon _{k}} \int
\frac{\di^{3} \krm '}{2\varepsilon _{k'}}\delta ^{4}(p-(k+k')/2)
\e ^{-\ii x \cdot (k'-k)} u_{\tau '}(k')_{a} \bar{u}_{\tau }(k)_{b}
\widehat{a}^{\dagger }_{h}(k)_{\tau }\widehat{a}_{h}(k')_{\tau '},
\end{equation}
while for the axial current the equation {\eqref{predominant}} is employed.
From now on we omit the subscript $h$ as only one hadronic species is involved.

It turns out that the correlator $\Delta W_{+}(x,p)_{ab}$ in the eq.~{\eqref{correlat}} 
involves the thermal
expectation values between four creation and annihilation operators
where the first two operators come from the Wigner operator in the eq.~
{\eqref{wigoppart}} and the remaining two operators from the
axial current operator in the eq.~{\eqref{predominant}}. Thanks
to the thermal Wick theorem, a four-operator thermal expectation value
can be reduced to the product of two-operator thermal expectation values
as follows:
\begin{equation*}
\mean{\widehat{a}^\dagger _{1} \widehat{a}_{2} \widehat{a}^\dagger _{3} \widehat{a}_{4}}_{c}=
\mean{\widehat{a}^\dagger _{1} \widehat{a}_{2} \widehat{a}^\dagger _{3} \widehat{a}_{4}}
-\mean{\widehat{a}^\dagger _{1} \widehat{a}_{2}}
\mean{ \widehat{a}^\dagger _{3} \widehat{a}_{4}}
=\mean{\widehat{a}^\dagger _{1} \widehat{a}_{4}}
\mean{ \widehat{a}_{2}\widehat{a}^\dagger _{3} } \; .
\end{equation*}
The two-operator thermal expectation values for non-interacting fields with 
the homogeneous grand-canonical ensemble operator $\wrho _{0}$ are given by:
\begin{equation}
\label{aadagger}
\begin{split} \mean{\widehat{a}^\dagger _\tau (k) \widehat{a}_{\sigma}(q)}_{
\beta (x)}=& \delta _{\tau \sigma }2\varepsilon _{q} \delta ^{3}(
\vec{k}-\vec{q})n\ped{F}(k,x),
\\
\mean{\widehat{a}_{\tau '}(k')\widehat{a}^\dagger _{\sigma '}(q')}_{
\beta (x)}=& \delta _{\tau '\sigma '}2\varepsilon _{q'}\delta ^{3}(
\vec{k\boldsymbol{'}}-\vec{q\boldsymbol{'}})(1-n\ped{F}(k',x)),
\end{split}
\end{equation}
where $n\ped{F}$ is the covariant Fermi-Dirac distribution function
\begin{equation*}
n\ped{F}(k,x) = \frac{1}{\e ^{\beta (x)\cdot k } + 1}.
\end{equation*}
All other combinations have vanishing expectation values.

By using the {\eqref{aadagger}}, after some simple calculation, both terms
on the right hand side of the equation {\eqref{eq:Linear}} can be worked
out:
\begin{equation}
\label{wigner0}
\mean{\wW_{+}(x,p)}_{\beta (x)}=\left (m+\gamma ^{\mu }p_{\mu
}\right ) \delta (p^{2}-m^{2})\theta (p_{0}) \frac{1}{(2\pi )^{3}} n
\ped{F}\left ( p\right ),
\end{equation}
and:
\begin{align} \label{wigner1}
\Delta W(x,p)_{+ab} = & \int _{\Sigma } \di \Sigma_{\rho}(y) \; \zeta _{A}(y) 
\int _{0}^{1}    \di  z \; \frac{1}{(2\pi )^{6}} \int \frac{\di^{3} \krm \, 
\di^{3} \krm '}{4\varepsilon _{k} \varepsilon _{k'}} \; 
\delta ^{4}\left (p-\frac{k+k'}{2} \right ) n\ped{F}(k,x)(1-n\ped{F}(k',x)) \\ \nonumber
&\times {\mathcal{A}}^{\rho }(k,k')_{ab}\, \e ^{\ii (k-k')\cdot (x-y)} \e^{z(k-k')\cdot \beta (x)}
\end{align}
where we defined:
\begin{equation*}
{\mathcal{A}}^{\rho }(k,k') \equiv (\slashed{k}'+m) \left [G_{A1}\left (Q^{2}
\right ) \gamma ^{\rho }\gamma ^{5} + \frac{{k'}^{\rho }-k^{\rho }}{2m} G_{A2}
\left (Q^{2}\right ) \gamma ^{5}\right ] (\slashed{k}+m) ,
\end{equation*}
where now $Q=(k'-k)$ because of the {\eqref{aadagger}}, and use has been
made of the known relation:
\begin{equation*}
\sum _{\sigma }u_{\sigma }(k) \bar{u}_{\sigma }(k)=\slashed{k}+m.
\end{equation*}

We can now work out an approximated expression of the mean spin vector
due to the axial chemical potential. By replacing the Wigner function in
the eq.~{\eqref{SM_eq:PolS}} with its local equilibrium approximation {\eqref{eq:Linear}}, 
and making use of the {\eqref{wigner0}} taking into account
the known traces of the $\gamma $ matrices, we are left with:
\begin{equation}
\label{spinchi1}
S^{\mu }_{\chi }(p)=\frac{1}{2}
\frac{\int _{\Sigma }\di  \Sigma \cdot p \;
\tr \left [\gamma ^{\mu }\gamma ^{5} \Delta W_{+}(x,p)\right ]}{\int _{\Sigma }\di \Sigma \cdot p 
\; \tr \left [\mean{\widehat{W}_{+}(x,p)}_{\beta (x)}+\Delta W_{+}(x,p)\right ]}
\end{equation}
as the term due to eq.~{\eqref{wigner0}} in the numerator gives vanishing
contribution. To proceed, we need to calculate some traces:
\begin{equation*}
\begin{split} \tr &\left (\slashed{p}+m\right ) = 4m
\\
\tr &\left [(\slashed{k}'+m)\gamma ^{\rho }\gamma ^{5}(\slashed{k}+m)
\right ]=0,
\\
\tr &\left [(\slashed{k}'+m)\gamma ^{5}(\slashed{k}+m)\right ]=0,
\\
\tr &\left [\gamma ^{\mu }\gamma ^{5}(\slashed{k}'+m)\gamma ^{5}(
\slashed{k}+m)\right ]= -4m({k'}^{\mu }-k^{\mu })
\\
\tr &\left [\gamma ^{\mu }\gamma ^{5}(\slashed{k}'+m)\gamma ^{\rho
}\gamma ^{5}(\slashed{k}+m)\right ]= -4\Big (\eta ^{\mu \rho }(m^{2}+k
\cdot k')-k^{\rho }{k'}^{\mu }-k^{\mu }{k'}^{\rho }\Big ).
\end{split}
\end{equation*}
By plugging the equations {\eqref{wigner0}} and {\eqref{wigner1}} into the 
{\eqref{spinchi1}} and using the above trace formulae, the following expression
is found for the mean spin vector:
\begin{equation}
\label{SM_eq:S_step1}
\begin{split} S^{\mu }_{\chi }(p)=&-\frac{2}{\mathcal{D}} \int _{\Sigma }
\di  \Sigma(x)\cdot p \int _{\Sigma }    \di  \Sigma _{\rho }(y) \; \zeta _{A}(y)
\int _{0}^{1} \frac{\di  z}{(2\pi )^{6}}
\int \frac{\di^{3} \krm }{2\varepsilon _{k}}\int
\frac{\di^{3} \krm '}{2\varepsilon _{k'}} \delta ^{4}\left (p-
\frac{k+k'}{2} \right )
\\
& \times {\mathcal{B}}^{\mu \rho }(k,k') n\ped{F}(k,x)(1-n\ped{F}(k',x))
\e ^{\ii (k-k')\cdot (x-y)} \e ^{z(k-k')\cdot \beta (x)},
\end{split}
\end{equation}
where:
\begin{equation}
\label{SM_eq:MomentaS}
{\mathcal{B}}^{\mu \rho }(k,k') \equiv G_{A1}(Q^{2})\left [\eta ^{\mu \rho }(m^{2}+k
\cdot k') -k^{\rho }{k'}^{\mu }-k^{\mu }{k'}^{\rho }\right ] +\frac{1}{2} G_{A2}
\left (Q^{2}\right ) ({k'}^{\mu }-k^{\mu }) ({k'}^{\rho }-k^{\rho })
\end{equation}
and ${\mathcal{D}}$ is the denominator in the leading order approximation:
\begin{equation}
\label{SM_eq:Den}
{\mathcal{D}}=\frac{4m}{(2\pi )^{3}}\int _{\Sigma }    \di  \Sigma
\cdot p \; \delta (p^{2}-m^{2})\theta (p_{0}) n\ped{F}(p).
\end{equation}

The {\eqref{SM_eq:S_step1}} is a double integral in $x,y$ which can be recast
as:
\begin{equation*}
S^{\mu }_{\chi }(p)=-\frac{2}{\mathcal{D}}\int _{\Sigma } \di\Sigma (x) \cdot p 
\; \int _{\Sigma } \di \Sigma _{\rho }(y) \zeta_{A}(y) \; G^{\mu\rho}(\beta (x),x-y)
\end{equation*}
where the function $G$ results from the integration in
$\krm ,\krm ^{\prime },z$. The function $G$ decays on microscopic length
scales as a function of its argument $x-y$ whereas the function
$\zeta _{A}$ supposedly varies significantly over a longer length scale,
in the hydrodynamic picture. Therefore, one can obtain a good approximation
of the above expression by replacing $\zeta _{A}(y)$ with
$\zeta _{A}(x)$ and taking it out of the $y$ integral. By doing so, only
an exponential is left to be integrated in $y$ in the eq.~{\eqref{SM_eq:S_step1}}:
\begin{equation*}
\int _{\Sigma }    \di  \Sigma _{\rho }(y) \; \zeta _{A}(y) \e ^{
\ii (k-k')\cdot (x-y)} \simeq \zeta _{A}(x)\int _{\Sigma }
\di  \Sigma _{\rho }(y) \; \e ^{\ii (k-k') \cdot (x-y)}.
\end{equation*}
To evaluate the integral over the hypersurface $\Sigma $, one can take
advantage of the Gauss theorem. By denoting with $\Omega \ped{B}$ the space-time
region encompassed by the 3D hypersurfaces $\Sigma _{\mathrm{FO}}$ and
$\Sigma _{B}$ which is the hyperplane region connecting the
$\Sigma \ped{FO}$ boundaries (see {Fig.~\ref{fig:fo}}):
\begin{equation*}
\int _{\Sigma }    \di  \Sigma _{\rho }(y)\; \e ^{\ii (k-k')
\cdot (x-y)} = \int _{\sigma _{\pm }}     \di  \Sigma _{\rho }(y)\;
\e ^{\ii (k-k')\cdot (x-y)} + \int _{\Sigma \ped{B}}
  \di  \Sigma _{\rho }(y)\; \e ^{\ii (k-k')\cdot (x-y)}   -
  \ii (k  -  k')_{\rho }\int _{\Omega \ped{B}} \di^{4} y\, \e ^{\ii (k-k')\cdot (x-y)}.
\end{equation*}
The contribution afrom the hyperbolic branches $\sigma _{\pm }$, which have
not even entered the plasma phase (see {Fig.~\ref{fig:fo}}), can be neglected
altogether, especially at high energy. The 3D hypersurface
$\Sigma \ped{B}$ is a subset of a hyperplane parallel to $t=0$ in the center-of-mass
frame (see {Fig.~\ref{fig:fo}}), thus
$\di  \Sigma _{\rho }= \hat{t}_{\rho }\di^{3} {\mathrm{y}} = \delta ^{0}_{\rho }
\di^{3} {\mathrm{y}}$. If it is large enough, one can approximate
it with a Dirac $\delta $:
\begin{equation*}
\int _{\Sigma \ped{B}}    \di  \Sigma _{\rho }(y)\, \e ^{
\ii (k-k') \cdot (x-y)}= \hat{t}_{\rho }\int \di^{3} y\,
\e ^{\ii (k-k')\cdot (x-y)}
\simeq \hat{t}_{\rho }(2\pi )^{3}\delta ^{3}(\vec{k}-\vec{k\boldsymbol{'}}).
\end{equation*}
Likewise, in the same approximation, the integral over the region
$\Omega \ped{B}$ multiplied by $(k-k')$ vanishes and one is finally left
with the approximation:
\begin{equation}
\label{eq:ApproxAxial}
\int _{\Sigma }    \di  \Sigma _{\rho }(y)\zeta _{A}(y) \e ^{
\ii (k-k')\cdot (x-y)} \simeq \zeta _{A}(x)\hat{t}_{\rho }(2\pi )^{3}
\delta ^{3}(\vec{k}-\vec{k\boldsymbol{'}}).
\end{equation}

With ${\mathbf{k}} = {\mathbf{k}}'$, being $k$ on-shell, we have $k=k'$ and
$Q=(k'-k)=0$. Therefore, the equation~{\eqref{SM_eq:MomentaS}} simplifies
to:
\begin{equation*}
{\mathcal{B}}^{\mu \rho }(k,k)=2 g_{h} \left (\eta ^{\mu \rho }m^{2} - k^{\rho }k^{\mu }\right ),
\end{equation*}
where $g_{h}=G_{A1}(0)$ is the axial charge, that is the matrix element~{\eqref{jmatel2}}
at zero momentum transfer. With the approximation~{\eqref{eq:ApproxAxial}}
we can readily integrate the expression~{\eqref{SM_eq:S_step1}} in
$k'$ and we obtain
\begin{equation*}
S^{\mu }_{\chi }(p) \simeq -\frac{2g_{h}}{\mathcal{D}}\int _{\Sigma }
\di  \Sigma (x)\cdot p \; \zeta _{A}(x) \int _{0}^{1}
\frac{\di  z}{(2\pi )^{3}} \int
\frac{\di^{3} \krm }{2\varepsilon _{k}}
\frac{1}{2\varepsilon _{k}} \delta ^{4}\left (p-k \right ) 2\left [
\hat{t}^{\mu }m^{2} -\varepsilon _{k} k^{\mu }\right ] n\ped{F}(k,x)(1-n
\ped{F}(k,x)).
\end{equation*}
Now, the dependence on $z$ is gone and the integration in $z$ is thus trivial.
Moreover:
\begin{equation*}
\int     \frac{\di^{3} \krm }{2\varepsilon _{k}}\delta ^{4}(p-k)f^\mu(k)=
    \int     \di^{4} k \; \delta (k^{2}    -m^{2})\theta (k_{0})
\delta^{4}(k  -  p)f^\mu(k) =\theta (p_{0})\delta (p^{2}-m^{2})f^\mu(p),
\end{equation*}
where
\begin{equation*}
f^\mu(k)=
\frac{\hat{t} ^{\mu }m^{2} -\varepsilon _{k} k^{\mu }}{\varepsilon _{k}}
n\ped{F}(k,x)(1-n\ped{F}(k,x)).
\end{equation*}
By using the previous results and replacing the denominator~{\eqref{SM_eq:Den}},
the final expression of the mean spin vector, at the leading order in the
axial chemical potential, is obtained:
\begin{equation*}
S^{\mu }_{\chi }(p)=\frac{g_{h}}{2}
\frac{\int _{\Sigma }\di  \Sigma (x) \cdot p \; \zeta _{A}(x)
n\ped{F}(p,x) \left (1-n\ped{F}(p,x) \right ) \delta (p^{2}-m^{2})\theta (p_{0})}
{\int _{\Sigma }\di  \Sigma (x) \cdot p \; n\ped{F}(p,x) \delta (p^{2}-m^{2})\theta (p_{0})}
\frac{\varepsilon p^{\mu }- m^{2}\hat{t}^{\mu }}{m\varepsilon }.
\end{equation*}
Since the integration over the hypersurface puts the momentum $p$ on-shell~\cite{Becattini:2020sww},
the delta functions $\delta (p^{2}-m^{2})$ give rise to an infinite constant
and cancel out in the ratio, while $\theta (p_{0})$ becomes redundant.
Therefore, the mean spin vector induced by chiral imbalance, at the leading
order in the axial chemical potential, is:
\begin{equation*}
\begin{split} S^{\mu }_{\chi }(p)=&\frac{g_{h}}{2}
\frac{\int _{\Sigma }\di  \Sigma \cdot p \;
\zeta _{A} n\ped{F}\left (1-n\ped{F}\right )}{\int _{\Sigma }\di  \Sigma \cdot p \;
n\ped{F}} \frac{\varepsilon p^{\mu }- m^{2}\hat{t}^{\mu }}{m\varepsilon }
\end{split}
\end{equation*}
where the arguments have been omitted.

\end{document}